\documentclass[12pt]{article}
\usepackage{amssymb,latexsym,amsmath,amsthm,bm}
\usepackage{authblk}

\theoremstyle{definition}

\usepackage[dvips]{color}
\usepackage{graphicx}
\usepackage[mathscr]{eucal}
\usepackage{booktabs}
\graphicspath{{fig/}}

\begin{document}
\baselineskip=0.3377in
\def\appendix{\par \setcounter{section}{0} \def\thesection{APPENDIX:}}
\newdimen\jot \jot=2mm

\renewcommand{\baselinestretch}{1.5}

\newcommand{\1}{{1}}\newcommand{\0}{{0}}
\newcommand{\I}{{I}}

\newcommand{\fb}{{f}}
\newcommand{\mb}{{m}}
\newcommand{\Jb}{{J}}
\newcommand{\ub}{{u}}
\newcommand{\rmspan}{{\rm span}}
\newcommand{\rmvec}{{\rm vec}}
\newcommand{\rmdim}{{\rm dim}}
\newcommand{\rmcov}{{\rm cov}}
\newcommand{\rmE}{{\rm E}}
\newcommand{\rmacov}{{\rm acov}}
\newcommand{\rmdiag}{{\rm diag}}
\newcommand{\rmeig}{{\rm Eig}}

\newtheorem{thm}{Theorem}
\newtheorem{rmk}{Remark}
\newtheorem{lem}{Lemma}
\newtheorem{prop}{Proposition}
\newtheorem{propty}{Property}
\newtheorem{eg}{Example}
\newtheorem{defn}{Definition}
\newtheorem{cor}{Corollary}
\newcommand{\indep}{\;\, \rule[0em]{.03em}{.67em} \hspace{-.25em}
\rule[0em]{.65em}{.03em} \hspace{-.25em}
\rule[0em]{.03em}{.67em}\;\,}

\title{A Two-Stage Dimension Reduction Method for Induced Responses and Its Applications}
\author{Hung Hung\\
Institute of Epidemiology and Preventive Medicine, \\National Taiwan University, Taipei, Taiwan, R.O.C.\\
hhung@ntu.edu.tw}
\date{}
\maketitle

\begin{abstract}

Researchers in the biological sciences nowadays often encounter the
curse of high-dimensionality, which many previously developed
statistical models fail to overcome. To tackle this problem,
sufficient dimension reduction aims to estimate the central subspace
(CS), in which all the necessary information supplied by the
covariates regarding the response of interest is contained.
Subsequent statistical analysis can then be made in a
lower-dimensional space while preserving relevant information.
Oftentimes studies are interested in a certain transformation of the
response (the induced response), instead of the original one, whose
corresponding CS may vary. When estimating the CS of the induced
response, existing dimension reduction methods may, however, suffer
the problem of inefficiency. In this article, we propose a more
efficient two-stage estimation procedure to estimate the CS of an
induced response. This approach is further extended to the case of
censored responses. An application for combining multiple biomarkers
is also illustrated. Simulation studies and two data examples
provide further evidence of the usefulness of the proposed method.\\

\end{abstract}

\section{Introduction}\label{sec.intro}

Consider the problem of inferring the association between the
response $Y$ and a $p$-dimensional vector of covariates $X$. Most
statistical methods perform well with a moderate size of $p$ in
comparison with the sample size. Unfortunately, we have trouble in
dealing with the problem when $p$ gets large, which is usually the
case in the biological sciences nowadays.
To improve statistical analysis, a preprocess is implemented first
to reduce the number of covariates and then the subsequent
statistical analysis is made based on those extracted covariates.
Sufficient dimension reduction aims to reduce the number of
covariates while preserving necessary information.
Specifically, it searches for a matrix $\Gamma\in
\textmd{R}^{p\times d}$ such that
\begin{eqnarray}
Y\indep X \mid \Gamma^TX, \label{dr}
\end{eqnarray}
where $\indep$ stands for statistical independence and $d\leq p$. An
equivalent statement is that the conditional distribution of $Y\mid
X$ and $Y\mid\Gamma^TX$ are the same. In other words, all the
information contained in $X$ regarding $Y$ can be obtained through
the lower-dimensional linear transformation $\Gamma^TX$.
Model (\ref{dr}) is very general without any extra specification for
the conditional distribution of $Y$ given $X$. It trivially holds
when $\Gamma$ is set to be the identity matrix and, hence, is useful
only when $d$ is adequately small.
Obviously,
it is $\rmspan(\Gamma)$ that is of interest to us, which is called
the dimension reduction subspace (Cook, 1994; Li, 1991) for the
regression of $Y$ on $X$. Under very general conditions, the
intersection of all such dimension reduction subspaces, denoted by
$\mathcal{S}_{Y\mid X}$, is still a dimension reduction subspace
(Cook, 1994) and is called the central subspace (CS). We thus assume
in the sequel the existence of $\mathcal{S}_{Y\mid
X}=\rmspan(\Gamma)$ with structural dimension
$\rmdim(\mathcal{S}_{Y\mid X})=d$. There have been many
methodologies proposed to estimate $\mathcal{S}_{Y\mid X}$,
beginning with the development of sliced inverse regression (SIR) of
Li (1991), including sliced average variance estimation (SAVE) of
Cook and Weisberg (1991), third-moment estimation of Yin and Cook
(2003), inverse regression (IR) of Cook and Ni (2005), directional
regression (DR) of Li and Wang (2007), discretization-expectation
estimation of Zhu et al. (2010), among others.

Oftentimes, researchers are interested in the induced response
$Y_g=g(Y)$ for a known function $g(\cdot)$ instead of the original
one. For example, the original response $Y$ in the Cardiac
Arrhythmia Study is a categorical random variable with value 1
referring to normal heart rhythm and values 2-16 for different types
of arrhythmia. In the phase of population screening, however, one
would merely like to distinguish patients with arrhythmia ($Y>1$)
from those without it ($Y=1$). In this case, $g(Y)=\I(Y\leq 1)$ is
of major interest, where $\I(\cdot)$ is the indicator function.
Taking the Angiography Cohort Study as another example, researchers
aim to predict a patient's 10-year vital status. In this study,
coronary artery disease (CAD)-related death time $Y$ is the original
response, and the induced response of interest is $g(Y)=\I(Y\leq
10)$. A far more complicated form of $g(\cdot)$ may, instead, be of
interest, depending on the nature of the study.

Similar to (\ref{dr}), there must exist for every $g(\cdot)$ a
$\Gamma_g\in \textmd{R}^{p\times d_g}$ such that
\begin{eqnarray}
Y_g\indep X \mid \Gamma_g^TX, \label{dr.c}
\end{eqnarray}
and one has the central subspace $\mathcal{S}_{Y_g\mid
X}=\rmspan(\Gamma_g)$ for the regression of $Y_g$ on $X$ with the
structural dimension $\rmdim(\mathcal{S}_{Y_g\mid X})=d_g$. We must
have $\mathcal{S}_{Y_g\mid X}\subseteq\mathcal{S}_{Y\mid X}$ since
$Y_g$ is a function of $Y$, but a more complicated inclusion
structure could exist. The following three examples demonstrate
various relationships between $\mathcal{S}_{Y|X}$ and
$\mathcal{S}_{Y_g|X}$ with $Y_g=\I(Y\leq t)$.

\noindent {\textbf{Example~1.}} Assume the conditional distribution
of $Y$ given $X$ is
\begin{eqnarray}
Y \mid X \sim \text{Gamma}(2\exp(\alpha^TX),~0.5) \label{ex.1}
\end{eqnarray}
which satisfies (\ref{dr}) with $\Gamma=\alpha$. It is easy to show
that (\ref{dr.c}) also holds with $\Gamma_g=\alpha$. In this case,
$\mathcal{S}_{Y_g\mid X}=\mathcal{S}_{Y\mid X}$ for every $t$.

\noindent {\textbf{Example~2.}} Assume the conditional distribution
of $Y$ given $X$ is
\begin{eqnarray}
\log Y \mid X \sim N\left(-\alpha_1^TX/\alpha_2^TX,\,
(\alpha_2^TX)^{-2}\right) \label{ex.2}
\end{eqnarray}
which satisfies (\ref{dr}) with $\Gamma=[\alpha_1,\,\alpha_2]$.
Provided $\alpha_2^TX>0$, $\text{pr}(Y_g=1\mid X)$ is a function of
$\Gamma_g^TX$, which satisfies (\ref{dr.c}) with
$\Gamma_g=\alpha_1+(\log t)\alpha_2$. In this case,
$\mathcal{S}_{Y_g\mid X}\subsetneq \mathcal{S}_{Y\mid X}$ and the
direction of $\mathcal{S}_{Y_g\mid X}$ changes as $t$ varies.

\noindent {\textbf{Example~3.}} Let the conditional hazard function
of $Y$ given $X$ be of the form
\begin{eqnarray}
\lambda(y\mid X)=\I(y < \tau_1)\exp(\alpha_1^TX)+\I(\tau_1\leq y <
\tau_2)\exp(\alpha_2^TX)+\I(\tau_2\leq y )\exp(\alpha_3^TX)
\label{ex.3}
\end{eqnarray}
which satisfies (\ref{dr}) with
$\Gamma=[\alpha_1,\,\alpha_2,\alpha_3]$. Moreover,
$\text{pr}(Y_g=1\mid X)$ is a function of $\Gamma_g^TX$, which
satisfies (\ref{dr.c}) with $\Gamma_g=\left[\alpha_1,\,\I(t\geq
\tau_1)\alpha_2,\, \I(t\geq \tau_2)\alpha_3 \right]$. In this case,
$\mathcal{S}_{Y_g\mid X}\subseteq \mathcal{S}_{Y\mid X}$ and
$\mathcal{S}_{Y_g\mid X}$ expands up to $\mathcal{S}_{Y\mid X}$ as
$t$ increases (i.e., the dimension also changes).

These examples highlight the importance of $\mathcal{S}_{Y_g\mid
X}$, because both the dimension and direction of the CS of $Y_g$ may
be different from the original CS, i.e., $\mathcal{S}_{Y\mid X}$ may
contain redundant directions if we are interested in $Y_g$ only. If
we simply treat $(Y_g,X)$ as the observed data, any dimension
reduction method can be directly applied to estimate
$\mathcal{S}_{Y_g\mid X}$. From a statistical point of view,
however, $Y$ must contain more information than $Y_g$ does,
therefore this direct method may suffer the problem of inefficiency.
We use model~(\ref{ex.1}) to demonstrate the potential drawback of
the direct method. Set $\alpha=(1,2,0)^T$ and generate $X$ from
$N(\0_3,0.8\,I_3+0.2\,\1_3 \1_3^T)$, where $I_a$ represents the $a
\times a$ identity matrix, $\1_a$ and $\0_a$ are $a \times 1$
vectors of ones and zeroes. Since $\mathcal{S}_{Y_g\mid
X}=\mathcal{S}_{Y\mid X}$, SIR is implemented to estimate
$\rmspan(\Gamma_g)$ based on $(Y,X)$ and $(Y_g,X)$ separately with
$t=t_{50}$, where $t_a$ satisfies $\text{pr}(Y\leq t_a)=a\%$. The
first element of the estimates is always forced to be one since only
the direction is relevant. Simulation results with sample size 300
and 500 replications performed give the means and standard errors of
the estimates as $(1.000,2.030\pm0.261,0.003\pm0.115)^T$ under
$(Y_g,X)$, and $(1.000,1.995\pm0.071,0.001\pm0.030)^T$ under
$(Y,X)$. Although both methods can accurately estimate the true
direction $(1,2,0)^T$, the standard errors for SIR based on
$(Y_g,X)$ are larger. We detect even larger biases and errors for
other choices of $t$, especially for $t$ near the boundaries. The
main theme of this paper is thus to propose a more efficient
estimation procedure for $\mathcal{S}_{Y_g\mid X}$ based on $(Y,X)$.



\section{A Two-Stage Estimation Procedure}\label{sec.complete}

Some notation is introduced first. For a square matrix $A$, let
$\rmeig(A;a)$ be the function which maps $A$ into its $a$ leading
eigenvectors. The observed data $(Y_i,X_i)$ is a random copy of
$(Y,X)$. Following the setting of Cook and Ni (2005), we may assume
$Y$ has a finite support $\{1,\cdots,h\}$. In the case of a
continuous response, it can be categorized as suggested by Li
(1991). Let $Z=\Sigma^{-1/2}(X-\mu)$ be the standardized version of
$X$, where $\mu=\rmE[X]$ and $\Sigma=\rmcov(X)$. Owing to
$\Sigma^{-1/2}\mathcal{S}_{Y\mid Z}=\mathcal{S}_{Y\mid X}$ and
$\Sigma^{-1/2}\mathcal{S}_{Y_g\mid Z}=\mathcal{S}_{Y_g\mid X}$,
there is no difference in considering the dimension reduction
problem under $Z$-scale. In this section, we will consider the
estimation of $B$ and $B_g$, the basis of $\mathcal{S}_{Y\mid Z}$
and $\mathcal{S}_{Y_g\mid Z}$, respectively, and transform back to
the original scale via $\Gamma=\Sigma^{-1/2}B$ and
$\Gamma_g=\Sigma^{-1/2}B_g$. In practice, $Z$ is replaced with $\hat
Z=\hat\Sigma^{-1/2}(X-\hat\mu)$ by plugging in the usual moment
estimators $\hat \mu$ and $\hat \Sigma$. The structural dimensions
$d$ and $d_g$ are assumed to be already known. The selection of
$(d,d_g)$ will be discussed later.

We start by reviewing a general estimation procedure for
$\mathcal{S}_{Y\mid Z}$. Most dimension reduction methods aim to
construct a symmetric kernel matrix $K$ (if $K$ is not symmetric,
$KK^T$ is used instead) based on $(Y,X)$ satisfying the property
\begin{eqnarray}
\rmspan(K)=\mathcal{S}_{Y\mid Z}. \label{ker.m}
\end{eqnarray}
A basis of $\mathcal{S}_{Y\mid Z}$ is then given by $B=\rmeig(K;d)$.
At the sampling level, $B$ is estimated by $\hat B=\rmeig(\hat
K;d)$, where $\hat K$ is a sample analogue of $K$.
For example, SIR considers
\begin{eqnarray}
K_{\rm SIR}=\Sigma^{-1/2}M\Sigma^{-1/2},~~~M=\rmcov(\rmE[X\mid
Y])=(\mb-\mu\1_h^T) D_\fb (\mb-\mu\1_h^T)^T,\label{ker.sir}
\end{eqnarray}
where $\mb=[m_1,\cdots,m_h]$ with $m_i=\rmE[X\mid Y=i]$,
$\fb=(f_1,\cdots,f_h)$ with $f_i=\text{pr}(Y=i)$, and
$D_\fb=\text{diag}(\fb)$. A sample analogue $\hat K_{\rm SIR}$ is
obtained by plugging the moment estimators $\hat \mb$, $\hat \fb$,
$\hat \mu$, and $\hat \Sigma$ into $K_{\rm SIR}$. It should be noted
that property (\ref{ker.m}) does not hold without any cost.
Depending on the choice of $K$, different conditions are imposed to
ensure its validity. Inverse regression methods, such as SIR,
commonly assume the linearity condition (\textbf{(A1)}: $\rmE[Z\mid
A^TZ]$~is a linear function of $Z$ for any matrix $A$), which is
equivalent to assuming the ellipticity of $X$ (Eaton, 1986).

Turning to the estimation of $\mathcal{S}_{Y_g\mid Z}$ for any given
$g(\cdot)$, parallel to (\ref{ker.m}), based on $(Y_g,X)$ we find
the symmetric kernel matrix $K_g$ satisfying
\begin{eqnarray}
\rmspan(K_g)=\mathcal{S}_{Y_g\mid Z}, \label{ker.mg}
\end{eqnarray}
and the basis of $\mathcal{S}_{Y_g\mid Z}$ which is of major
interest is defined to be $B_g=\rmeig(K_g;d_g)$.
The direct estimation method then substitutes an estimator $\hat
K_g$ for $K_g$, and estimates $B_g$ by $\rmeig(\hat K_g;d_g)$.
Similar to (\ref{ker.sir}), $K_g$ of SIR is given by
\begin{eqnarray}
K_{g,\rm SIR}=\Sigma^{-1/2}M_g\Sigma^{-1/2},~~~M_g=\rmcov(\rmE[X\mid
Y_g])=(\mb_g-\mu\1_s^T) D_{\fb_g}
(\mb_g-\mu\1_s^T)^T,\label{kerg.sir}
\end{eqnarray}
where $\mb_g=[m_{g1},\cdots,m_{gs}]$, $m_{gi}=\rmE[X \mid Y_g=i]$,
$\fb_g=(f_{g1},\cdots,f_{gs})$, $f_{gi}=\text{pr}(Y_g=i)$, and $s$
is the number of categories of $Y_g$. Note that $s\leq h$ since
$Y_g$ is a function of $Y$. The sample analogue $\hat K_{g,\rm SIR}$
can be obtained by plugging the moment estimators $\hat \mb_g$,
$\hat \fb_g$, $\hat \mu$, and $\hat \Sigma$ into $K_{g,\rm SIR}$. We
have seen in the end of Section~\ref{sec.intro} that direct
estimation based on $(Y_g,X)$ may lose information, and we attempt
to propose a more efficient estimation procedure. First observe that
under the validity of (\ref{ker.m}) and (\ref{ker.mg}), we must have
\begin{eqnarray}\label{relationship}
K_g=P_BK_gP_B,
\end{eqnarray}
where $P_B=BB^T$ is the orthogonal projection matrix onto
$\rmspan(B)$. Although (\ref{relationship}) is straightforward, it
motivates us to estimate $K_g$ by $\hat P_B\hat K_g\hat P_B$, where
$\hat P_B=\hat B\hat B^T$ is an estimate of $P_B$. It is the
projection $\hat P_B$ that utilizes the extra information in
$(Y,X)$, and results in an expected gain in efficiency.
Details of the procedure are listed below:
\begin{itemize}
\item[1.] Based on $(Y,X)$, apply a dimension reduction method to obtain $\hat K$
and, hence, $\hat P_B$.

\item[2.] Based on $(Y_g,X)$, apply a dimension reduction method to obtain $\hat
K_g$.

\item[3.] Estimate $B_g$ by $\hat B_g=\rmeig(\hat P_B\hat K_g\hat
P_B;d_g)$.
\end{itemize}
With $\hat B_g$ obtained, we then estimate a basis of
$\mathcal{S}_{Y_g|X}$, say $\Gamma_g$, by
$\hat\Gamma_g=\hat\Sigma^{-1/2}\hat B_g$.
The $n^{1/2}$-consistency of $\hat \Gamma_g$ is a direct consequence
provided $\hat K$ and $\hat K_g$ are also $n^{1/2}$-consistent. We
call the two-stage estimation procedure ``A-B'' hereafter, if method
A is used in Step~1 and method B in Step~2.
As SIR is the most widely applied dimension reduction method, the
following theorem, which guarantees that SIR-SIR is more efficient
than SIR, highlights the desirability of using our two-stage
estimation procedure. We use ``acov'' to denote the asymptotic
covariance, and $A\geq0$ to indicate $A$ is positive semi-definite.
The proof is deferred to the Appendix.

\begin{thm}\label{thm.1}
Let $\hat \Gamma_g$ be obtained from SIR-SIR, and let
$\tilde\Gamma_g=\hat\Sigma^{-1/2}\rmeig(\hat K_{g,\rm SIR};d_g)$ be
the direct estimate of $\Gamma_g$ from SIR. In addition to the
linearity condition \textbf{(A1)} above, assume the validity of
\text{\textbf{(A2)}}: $\rmcov(\nu^TZ\mid B^TZ)$ is non-random for
any $\nu\bot~\mathcal{S}_{Y\mid Z}$. Then,
\begin{eqnarray}
\Delta=\rmacov\left(n^{1/2}\rmvec(\tilde
\Gamma_g-\Gamma_g)\right)-\rmacov\left(n^{1/2}\rmvec(\hat
\Gamma_g-\Gamma_g)\right)\geq 0.\nonumber
\end{eqnarray}
The equality holds if and only if $\rmspan(K_{g,\rm
SIR})\bigcap~\rmspan(K_{\rm SIR}-K_{g,\rm SIR})=\{0\}$, where
$K_{\rm SIR}$ and $K_{g,\rm SIR}$ are defined in (\ref{ker.sir}) and
(\ref{kerg.sir}).
\end{thm}

In the establishment of Theorem~\ref{thm.1}, in addition to the
linearity condition we require $\rmcov(\nu^TZ\mid B^TZ)$ to be
non-random for any $\nu$ in the complement of $\mathcal{S}_{Y\mid
Z}$. These conditions are not that restrictive and can be generally
satisfied. As argued by Li and Wang (2007), \textbf{(A1)-(A2)} are
shown to approximately hold when $p$ is large. Moreover,
\textbf{(A2)} is valid when $X$ is normally distributed. Although
normality is a stronger condition, it can be approximated by making
a power transformation of $X$. One implication of
Theorem~\ref{thm.1} is that the total asymptotic variance of
$\tilde\Gamma_g$ is strictly larger than that of $\hat \Gamma_g$
provided $\Delta\neq 0$. The only possibility of no efficiency gain
(i.e., $\Delta=0$) is when $\rmspan(K_{g,\rm SIR})$ and
$\rmspan(K_{\rm SIR}-K_{g,\rm SIR})$ have no common element except
the zero point. This is reasonable since, under this situation, all
the information about $S_{Y_g|Z}$ contained in $K_{\rm SIR}$ resides
in $K_{g,\rm SIR}$ and knowing the ``residual'' $(K_{\rm
SIR}-K_{g,\rm SIR})$ contributes nothing to the construction of
$S_{Y_g|Z}$. Hence, we will gain nothing from SIR-SIR. A formal test
for this condition is beyond the scope of this article and will be
investigated in a future study. In summary, SIR-SIR is expected to
perform well in most of the situations except the rather restrictive
special case. This fact is also demonstrated by our simulation
studies in Section~4, where the efficiency gain of the two-stage
method is obviously detected.



The structural dimensions $d$ and $d_g$ should be determined before
practical implementation. To estimate $d$, most methods rely on a
sequence of hypothesis tests
(Li, 1991; Cook and Lee, 1999, Cook and
Yin, 2001). These methods, however, may not be readily applicable
for the selection of $d_g$. To simplify the estimation procedure, we
alternatively suggest two approaches to select $(d,d_g)$. One is to
adopt the maximal eigenvalue ratio criterion (MERC) proposed by Luo,
Wang, and Tsai (2009). Let $\hat\lambda_i$ be the eigenvalue of
$\hat K$ and define $\hat\rho_i=\hat\lambda_i/\hat\lambda_{i+1}$ for
$1\leq i\leq p-1$. It is proposed to select $d$ by $\hat
d=\arg\max_{1\leq i\leq d^*}\hat\rho_i$, where $d^*$ is a
pre-specified constant. The authors suggest using $d^*=5$ in
practice. Once $\hat d$ is obtained, we can estimate $d_g$ by a
similar procedure. Let $\hat\lambda_{g,i}$ be the eigenvalue of
$\hat P_B\hat K_g\hat P_B$ and define
$\hat\rho_{g,i}=\hat\lambda_{g,i}/\hat\lambda_{g,i+1}$ for $1\leq
i\leq \hat d-1$. Then $d_g$ is determined by $\hat
d_g=\arg\max_{1\leq i\leq \hat d-1}\hat\rho_{g,i}$. As to the second
method, note that the purpose of dimension reduction is to improve
regression or classification.
Thus, it is natural to select $(d,d_g)$ so that a measure of
classification accuracy is maximized. In Section~\ref{sec.data}
below, the classification accuracy obtained from cross-validation is
used in the Cardiac Arrhythmia Study, while the AUC (area under the
receiver operating characteristic (ROC) curve) is considered in the
Angiography Cohort Study to select $(d,d_g)$.

\begin{rmk}\label{rmk.save}
In our two motivating examples, $Y_g=\I(Y\leq t)$ is binary and,
hence, due to its nature, SIR can capture at most one direction of
$\mathcal{S}_{Y_g\mid Z}$. Alternatively, we can adopt SAVE in
Step~2. Cook and Lee (1999) showed that for a binary response, SAVE
is more comprehensive than SIR. The kernel matrix of SAVE is
\begin{eqnarray}\label{save}
K_{g,\rm
SAVE}=\left[\Sigma^{-1/2}(\mu_{t1}-\mu_{t0})~,\Sigma^{-1/2}(\Sigma_{t1}-\Sigma_{t0})\Sigma^{-1/2}\right]
\end{eqnarray}
with $\mu_{ti}=E[X\mid Y_g=i]$ and $\Sigma_{ti}=\rmcov(X\mid
Y_g=i)$, $i=0,1$. Its sample analogue $\hat K_{g,\rm SAVE}$ is
obtained by plugging moment estimators $\hat\mu_{t0}$,
$\hat\mu_{t1}$, $\hat\Sigma_{t0}$, $\hat\Sigma_{t1}$, and
$\hat\Sigma$ into (\ref{save}).
\end{rmk}

\section{Extension to Censored Response}\label{sec.censor}

Dimension reduction is usually applied in the field of life science
when the response of interest $Y$ represents the survival time of a
subject. An important issue in survival analysis is that the
response may be censored.
The exact survival time $Y$ (and hence $Y_g$) may not always be
observed and we can only observe $(Y^*,\delta,X)$ instead, where
$Y^*=\min\{Y,C\}$ is the last observed time, $\delta=\I(Y\leq C)$ is
the censoring status, and $C$ is the censoring time. Motivated from
two data examples in Section~1, our aim here is to modify SIR-SAVE
to estimate $\mathcal{S}_{Y_g\mid X}$ with the specific choice
$Y_g=\I(Y\leq t)$
under the validity of totally independent censorship $C \indep
(Y,X)$. The modified SIR-SIR will also be illustrated.
We note that totally independent censorship is satisfied in the
Angiography Cohort Study, since most of the patients are subject to
Type-I censoring.

Both SIR and SAVE in Steps~1-2 should therefore be modified. For SIR
in Step~1, observe that $\mathcal{S}_{(Y^*,\delta)\mid
Z}\subseteq\mathcal{S}_{(Y,C)\mid Z}=\mathcal{S}_{Y\mid Z}$,
where the first inclusion property holds since $(Y^*,\delta)$ is a
function of $(Y,C)$, and the last equality is true by the totally
independent censorship assumption.
Thus, we suggest using the modified kernel matrix
\begin{eqnarray*}
K_{\rm SIR}^*=\Sigma^{-1/2}M^*\Sigma^{-1/2},~~~M^*=\rmcov(\rmE[X\mid
Y^*,\delta])=(\mb^*-\mu\1_{h_0+h_1}^T) D_{\fb^*}
(\mb^*-\mu\1_{h_0+h_1}^T)^T, \label{c.ker.sir}
\end{eqnarray*}
where
$\mb^*=[m_{(0,1)}^*,\cdots,m_{(0,h_0)}^*,m_{(1,1)}^*,\cdots,m_{(1,h_1)}^*]$
with $m_{(i,j)}^*=\rmE[X\mid \delta=i,Y^*=j]$,
$\fb^*=(f_{(0,1)}^*,\cdots,f_{(0,h_0)}^*,f_{(1,1)}^*,\cdots,f_{(1,h_1)}^*)^T$
with $f_{(i,j)}^*=\text{pr}(\delta=i,Y^*=j)$, and $h_0\leq h$ and
$h_1\leq h$ denote the number of categories of $Y^*$ when $\delta=0$
and $\delta=1$.
Here the slice means, the $m_{(i,j)}^*$'s, are formed within those
patients with $\delta=0$ and $\delta=1$ separately. By plugging in
moment estimators $\hat\mb^*$, $\hat\fb^*$, $\hat \mu$, and
$\hat\Sigma$, the sample analogue $\hat K_{\rm SIR}^*$ is obtained.
This double slicing procedure was originally proposed by Li, Wang,
and Chen (1999), and
our point is to emphasize its validity under totally independent
censorship.

With regard to implementing SAVE in Step~2, we can still use the
kernel matrix $K_{g,\rm SAVE}$ in (\ref{save}) provided it can be
estimated based on $(Y^*,\delta,X)$. First observe that
\begin{equation}
E[X^{\otimes i}\mid Y_g=0]=-\frac{\int u^{\otimes i}
dS_{XY}(u,t)}{S_{XY}(-\infty,t)},~i=1,2, \label{mo.0}
\end{equation}
\begin{equation}
E[X^{\otimes i}\mid Y_g=1]=-\frac{\int u^{\otimes i}
d\{S_{XY}(u,-\infty)-S_{XY}(u,t)\}}{1-S_{XY}(-\infty,t)},~i=1,2,
\label{mo.1}
\end{equation}
where $a^{\otimes 1}=a$ and $a^{\otimes 2}=aa^T$ for a vector $a$,
and $S_{XY}(x,y)={\rm pr}(X>x,Y>y)$.
Here ``$>$'' is interpreted as component-wise for a vector. It
implies the $\mu_{ti}$'s and $\Sigma_{ti}$'s in (\ref{save}) are
functionals of $S_{XY}(x,y)$. Campbell (1981) and Burke (1988) have
separately proposed two different estimators of $S_{XY}(x,y)$,
denoted by $\hat{S}^{(c)}_{XY}(x,y)$ and $\hat{S}^{(b)}_{XY}(x,y)$.
By plugging $\hat S_{XY}^{(c)}(x,y)$ into (\ref{mo.0}) and $\hat
S_{XY}^{(b)}(x,y)$ into (\ref{mo.1}), we can estimate $\mu_{ti}$'s
and $\Sigma_{ti}$'s by
\begin{eqnarray*}
\hat \mu_{t0}^* =\frac{\sum_{i=1}^n X_i
\I(Y_i^*>t)}{\sum_{i=1}^n\I(Y^*_i>t)},&&\hat
\Sigma_{t0}^*=\frac{\sum_{i=1}^n X_i^{\otimes
2} \I(Y_i^*>t)}{n\hat{S}_Y(t)}-\{\hat \mu_{t0}^*\}^{\otimes 2},\\
\hat \mu_{t1}^* =\frac{1}{n}\sum_{i=1}^n\frac{X_i\delta_i
\I(Y_i^*\leq t)}{\{1-\hat{S}_Y(t)\}\hat{S}_C(Y_i^*)},&& \hat
\Sigma_{t1}^*=\frac{1}{n}\sum_{i=1}^n\frac{X_i^{\otimes 2}\delta_i
\I(Y_i^*\leq t)}{\{1-\hat{S}_Y(t)\}\hat{S}_C(Y_i^*)}-\{\hat
\mu_{t1}^*\}^{\otimes 2},\label{}
\end{eqnarray*}
where $\hat{S}_Y(y)$ and $\hat{S}_{C}(y)$ are Kaplan-Meier
estimators of $\text{pr}(Y>y)$ and $\text{pr}(C>y)$. Finally, a
modified estimator of $K_{g,\rm SAVE}$ is given by
\begin{eqnarray*}
\hat K_{g,\rm
SAVE}^*=\left[\hat\Sigma^{-1/2}(\hat\mu_{t1}^*-\hat\mu_{t0}^*)~,
\hat\Sigma^{-1/2}(\hat\Sigma_{t1}^*-\hat\Sigma_{t0}^*)\hat\Sigma^{-1/2}\right].
\end{eqnarray*}
The modified SIR-SAVE is then proposed by using $\hat K_{\rm SIR}^*$
and $\hat K_{g,\rm SAVE}^*$ in Steps~1-2.

\begin{rmk}
For binary $Y_g$, Cook and Lee (1999) showed that the population
kernel matrix of SIR can be expressed as
$\Sigma^{-1/2}(\mu_{t1}-\mu_{t0})$. The modified SIR-SIR is then
proposed by using $\hat K_{g,\rm
SIR}^*=\hat\Sigma^{-1/2}(\hat\mu_{t1}^*-\hat\mu_{t0}^*)$ in Step~2.
\end{rmk}

\section{Simulation Studies}\label{sec.sim}

We use models~(\ref{ex.2})-(\ref{ex.3}) to evaluate the performance
of our two-stage estimation procedure under different combinations
of sample sizes $(n=50,100)$, number of covariates $(p=10,20)$, and
censoring rates (${\rm CR}=0\%,25\%$). With censored data, the
modified procedure is implemented instead. To measure the closeness
of two spaces with basis $A$ and $A'$, we adopt the Frobenius norm
$\text{tr}\{(P_{A}-P_{A'})(P_{A}-P_{A'})\}^{1/2}$, where $P_{A}$ is
the orthogonal projection matrix onto $\rmspan(A)$. Simulations are
repeated 500 times.


For model~(\ref{ex.2}), set $\alpha_1=(3,0.9,-1.5,\0_{p-3}^T)^T$ and
$\alpha_2=(3,4.5,6,\0_{p-3}^T)^T$. We independently generate $u$ and
$r$ from $N_p(0,I_p)$ and Beta$(1.8,0.3)$, and define
$X=\mu+\Sigma^{1/2}ru(u^Tu)^{-1/2}$ with
$\Sigma=0.8I_p+0.2\1_p\1_p^T$ and $\mu=(0,3,0,\0_{p-3}^T)^T$. This
ensures the ellipticity of $X$. For the censored case, $C$ is
generated from Gamma$(2,1.71)$ so that CR$=25\%$. Both SIR-SIR and
SIR are implemented at $t=t_{30}$, $t_{50}$, and $t_{70}$. As for
the case of model~(\ref{ex.3}), we set
$\alpha_1=(20,0,0,\0_{p-3}^T)^T$, $\alpha_2=(0,15,0,\0_{p-3}^T)^T$,
$\alpha_3=(0,0,10,\0_{p-3}^T)^T$, and $(\tau_1,\tau_2)=(\log 2,\log
8)$, generate $X$ from
$N_p(-0.2\cdot\1_p,D(0.8I_p+0.2\cdot\1_p\1_p^T)D)$
with $D={\rm diag}(2,1,1,\1_{p-3}^T)$, and generate $C$ from
Gamma(1,8) to produce CR$=25\%$. We implement SIR-SAVE and SAVE at
$t=t_{45}$, $t_{65}$, and $t_{75}$ so that $d_g=1$, 2, and 3.
Various choices of the slicing number were examined and produced a
similar result. We thus use $h=10$ for SIR-SIR and SIR-SAVE, and
$(h_0,h_1)=(5,10)$ for the modified methods.

Simulation results are provided in Table~1. Compared with the
standard setting $(n,p,{\rm CR})=(100,10,0\%)$, an overall
observation is that SIR-SIR and SIR-SAVE outperform SIR and SAVE,
even for the cases of smaller sample size $(n=50)$, of more
``noise'' covariates $(p=20)$, and of censored response (CR$=25\%$).
The magnitude of efficiency gain from SIR-SIR is roughly the same
for every $t$ in model~(\ref{ex.2}). Interestingly, the efficiency
gain from SIR-SAVE in model~(\ref{ex.3}) becomes greater for larger
$t$. One reason is that the structural dimension of
$\mathcal{S}_{Y_g|X}$ also increases as $t$ does. With more
directions needing to be estimated, more information is required to
recover $\mathcal{S}_{Y_g|X}$, and we gain more from the two-stage
estimation procedure. It has been found empirically that SAVE is
less efficient than SIR. Li and Zhu (2007) showed that SAVE will not
attain $n^{1/2}$-consistency in general, while SIR will, even if the
number of samples in each slice is only 2. By combining SIR and
SAVE, we expect an efficiency gain from SIR-SAVE as shown in this
simulation.

\section{Data Examples}\label{sec.data}

\subsection{The Angiography Cohort Study}

Detailed description of the data can be found in Lee et al. (2006).
Briefly speaking, for each of 1050 traceable patients, four
biomarkers (CRP, SAA, IL-6, and tHcy) and the CAD-related time of
death were recorded with the aim of using the combined biomarkers to
accurately predict a patient's $t$-year vital status, and thus the
induced response of interest is $Y_g=\I(Y\leq t)$.
Hung and Chiang (2010) analyzed this data, combining biomarkers via
the extended generalized linear model (EGLM): $P(Y\leq t\mid
X)=G(t,\beta_t^TX)$, where $\beta_t$ is a $p\times 1$ time-varying
coefficient vector and $G(\cdot,\cdot)$ is an unknown link function
which is monotone increasing in its two arguments.
Under EGLM, $\beta_t^TX$ is promised to be optimal in distinguishing
$\{Y\leq t\}$ from $\{Y>t\}$, in the sense that the time-dependent
ROC curve (Heagerty, Lumley, and Pepe, 2000) is the highest among
all functions of $X$.

The EGLM also satisfies (\ref{dr.c}) with $Y_g=\I(Y\leq t)$,
$\mathcal{S}_{Y_g|X}=\rmspan(\Gamma_g)=\rmspan(\beta_t)$, and
$d_g=1$. Thus, $\Gamma_g^TX$ is also the optimal biomarker since any
monotone transformation of $\beta_t^TX$ will have the same
time-dependent ROC curve. Given that a censoring mechanism is
involved in this study, the modified SIR-SIR is applied to obtain
$\hat\Gamma_g$ in order to combine the biomarkers. We enter the
transformed biomarker $X_i/\text{sd}(X_i)$ to perform our analysis.
The analysis results with $d=3$ and $(h_0,h_1)=(2,4)$ are found in
Table~2. We remind the reader that the choice of these tuning
parameters attains the maximum of the time-dependent AUC as
mentioned in Section~\ref{sec.complete}. The absolute coefficient of
CRP is smallest at the beginning
and increases as time goes by. SAA has a totally different behavior,
where it has a larger effect initially but seems to be diminishing
at 3500 days. Both IL-6 and tHcy are found to play important roles
in predicting patient's vital status over time.
Interestingly, CRP has a reverse effect as compared with the other
three biomarkers. Table~2 provides the time-dependent AUC of the
composite biomarkers $\hat\Gamma_g^TX$ at day $t$, denoted by
$\mathcal{A}_t$ (see equation (8) of Chiang and Hung, 2010). The
larger the $\mathcal{A}_t$ values, the higher prediction power
$\hat\Gamma_g^TX$ has. One can see that most of the $\mathcal{A}_t$
values are greater than 0.7, especially at the beginning of the
study.
We also calculated $\mathcal{A}_t^*$ values, the maximal
time-dependent AUC of the method developed in Hung and Chiang
(2010), and a similar pattern to that of the $\mathcal{A}_t$ values
was detected (note that $\mathcal{A}_t\leq\mathcal{A}_t^*$ will
always hold for every $t$). In summary, SIR-SIR is easy to implement
and achieves acceptable AUC values.

\subsection{The Cardiac Arrhythmia Study}

The study consisted of 452 patients, each with 279 covariates. The
response $Y\in\{1,\cdots,16\}$ is a categorical random variable,
where 1 refers to ``normal'' and 2-16 refer to different classes of
arrhythmia. See G\"{u}venir et al. (1997) for details.

\begin{figure}[h]
\includegraphics[height=12.6cm]{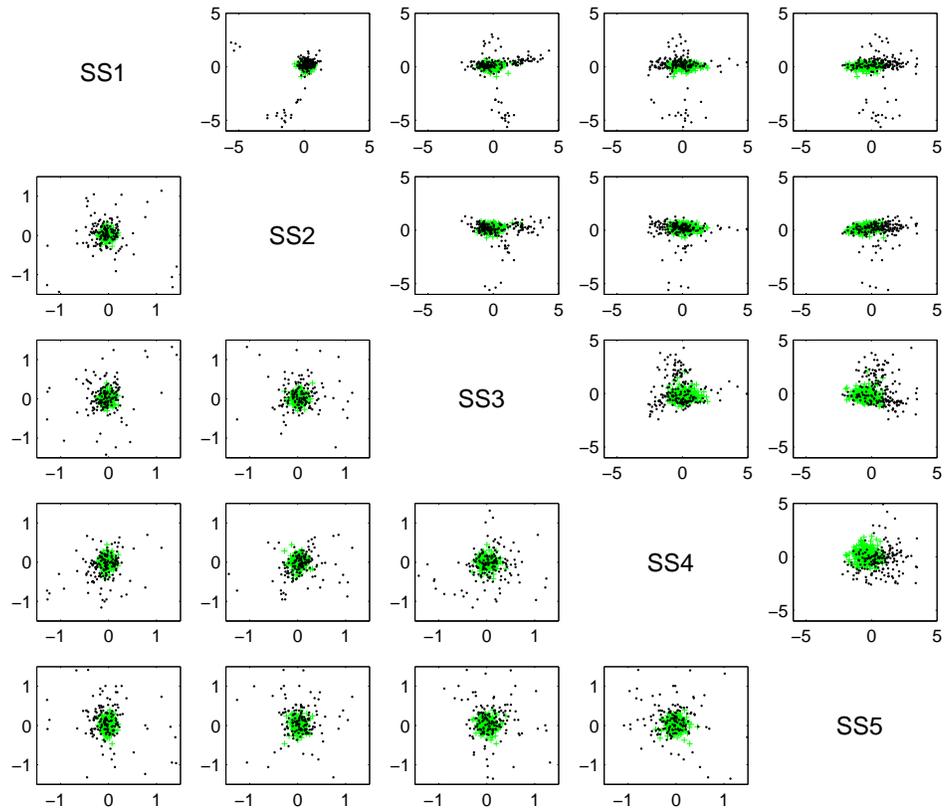}
\caption{\label{fg} The scatter plot matrix of extracted predictors
from SIR-SAVE (upper triangular panel) and SAVE (lower triangular
panel) with $(d,d_g)=(7,5)$. The green pluses and black dots
indicate the normal and abnormal patients.}
\end{figure}

To keep matters simple, we consider continuous predictors only and
use their first 100 principal components in our analysis. We are
interested in distinguishing normal patients $\{Y=1\}$ from abnormal
ones $\{Y>1\}$, i.e., $Y_g=\I(Y\leq 1)$. The scatterplots of the
extracted predictors (denoted by SS1,$\cdots$,SS5) from SIR-SAVE
with $(d,d_g)=(7,5)$ are provided in Figure~{\ref{fg}}.
Again, the selection of $(d,d_g)$ is such that the averaged
classification accuracy from cross-validation is maximized. It can
be seen that
SS1-SS3 demonstrate their ability to separate two groups via
variation, while SS4-SS5 attempt to separate two groups via
location. In every subplot, the normal group seems to have smaller
variation and locates in the center of a relatively large data cloud
of the abnormal group.
The bottom-left 10 subplots of Figure~{\ref{fg}} are scatterplots of
those extracted predictors taken from SAVE directly. It can be seen
that there is only a separation pattern of variation between the two
groups, but no obvious location difference. To further evaluate the
performance of those extracted predictors, we randomly separate the
data into a training set ($90\%$) and a test set ($10\%$), and then
implement quadratic discriminant analysis based on those extracted
predictors. The procedure with 200 replications gives SIR-SAVE the
averaged classification accuracy of $78\%$, while it is a mere
$70\%$ for SAVE.

\section{Discussion}\label{sec.discussion}

Although we have considered univariate responses only, there is
nothing different about carrying out the procedure with multivariate
responses, except that the kernel matrices $\hat K$ and $\hat K_g$
are constructed for multivariate responses $Y$ and $g(Y)$. A
multivariate response version of Theorem~\ref{thm.1} can be derived
with a proof analogous to the proof of the univariate case. We refer
to Li, Wen, and Zhu (2008) for some recent developments in dimension
reduction with multivariate responses. We note that the proposed
two-stage estimation procedure is a general framework, and is not
limited to any specific method. Depending on the purpose of a given
study, we may adopt any dimension reduction technique in either
Steps~1 or 2 of the procedure. Besides SIR-SIR and SIR-SAVE, we also
tested various combinations of SIR, SAVE, IR, and DR. Simulation
results (not shown here) all convey the same message that an
efficiency gain is significantly detected, which provides evidence
that the superiority of the two-stage procedure comes mainly from
using $\hat P_B\hat K_g\hat P_B$, and is not limited to any specific
choice of dimension reduction method.


\begin{center}
{\large REFERENCES}
\end{center}
\begin{description}
\item[] Anderson, W. N., Jr. and Duffin, R. J. (1969). Series and parallel addition of matrices.
\textit{J. Math. Anal. Appl.} \textbf{26}, 576-594.

\item[] Burke, M. D. (1988). Estimation of a bivariate distribution
function under random censorship. \textit{Biometrika} \textbf{75},
379-382.

\item[] Campbell, G. (1981). Nonparametric bivariate estimation with
randomly censored data. \textit{Biometrika} \textbf{68}, 417-423.

\item[] Chiang, C. T. and Hung, H. (2010). Nonparametric
estimation for time-dependent AUC. {\em J. Stat. Plan. Infer.}
\textbf{140}, 1162-1174.

\item[] Cook, R. D. and Weisberg, S. (1991). Discussion of ``Sliced inverse regression for dimension
reduction''. {\em J. Am. Stat. Assoc.} \textbf{86}, 328-332.

\item[] Cook, R. D. (1994). On the interpretation of regression plots.
{\em J. Am. Stat. Assoc.} \textbf{89}, 177-189.


\item[] Cook, R. D. and Lee, H. (1999). Dimension reduction in
binary response regression. {\em J. Am. Stat. Assoc.} \textbf{94},
1187-1200.

\item[] Cook, R. D. Yin, X. (2001). Dimension reduction and visualization in discriminant
analysis (with discussion). {\em Aust. Nz. J. Stat.} \textbf{43},
147-199.


\item[] Cook, R. D. and Ni, L. (2005). Sufficient dimension reduction via
inverse regression: a minimum discrepancy approach. {\em J. Am.
Stat. Assoc.} \textbf{100}, 410-427.


\item[] Eaton, M. L. (1986). A characterization of spherical distributions.
{\em J. Multivariate Anal.} \textbf{20}, 272-276.


\item[] G\"{u}venir, H. A., Acar, B., Demir\"{o}z, G, and \c{C}ekin, A. (1997). A supervised
machine learning algorithm for arrhythmian analysis. {\em Computers
in Cardiology} \textbf{24}, 433-436.


\item[] Heagerty, P. J., Lumley, T. and Pepe, M. (2000).
Time-dependent ROC curves for censored survival data and a
diagnostic marker. {\em Biometrics} \textbf{56}, 337-344.

\item[] Hung, H. and Chiang, C. T. (2010). Optimal composite
markers for time-dependent receiver operating characteristic curves
with censored survival data. {\em Scand. J. Stat.} \textbf{37},
664-679.

\item[] Lee, K. W. J., Hill, J. S., Walley, K. R., and Frohlich, J.
J. (2006). Relative value of multiple plasma biomarkers as risk
factors for coronary artery disease and death in an angiography
cohort. {\em Canadian Medical Association Journal} \textbf{174},
461-466.

\item[] Li, B. and Wang, S. (2007). On directional regression for dimension reduction.
{\em J. Am. Stat. Assoc.} \textbf{102}, 997-1008.

\item[] Li, B., Wen, S. and Zhu, L. (2008). On a projective resampling method
for dimension reduction with multivariate responses. {\em J. Am.
Stat. Assoc.} \textbf{103}, 1177-1186.

\item[] Li, K. C. (1991). Sliced inverse regression for dimension reduction (with discussion).
{\em J. Am. Stat. Assoc.} \textbf{86}, 316-342.

\item[] Li, K. C., Wang, J. L., and Chen, C. H. (1999). Dimension reduction for censored regression data.
{\em Ann. Stat.} \textbf{27}, 1-23.

\item[] Li, Y. X. and Zhu, L. X. (2007). Asymptotics for sliced average variance estimation. {\em
Ann. Stat.} \textbf{35}, 41-69.

\item[] Luo, R., Wang, H., and Tsai, C. L. (2009). Contour projected dimension reduction.
{\em Ann. Stat.} \textbf{37}, 3743-3778.

\item[] Saracco, J. (1997).
An asymptotic theory for sliced inverse regression. {\em Commun.
Stat. - Theor. M.} \textbf{26}, 2141-2171.

\item[] Tyler, D. E. (1981). Asymptotic inference for eigenvectors. {\em
Ann. Stat.} \textbf{9}, 725-736.

\item[] Yin, X. and Cook, R. D. (2003). Estimating the central subspaces via inverse third moments.
{\em Biometrika} \textbf{90}, 113-125.

\item[] Zhu, L., Wang, T., Zhu, L., and Ferr\'{e}, L. (2010).
Sufficient dimension reduction through discretization-expectation
estimation. {\em Biometrika} \textbf{97}, 295-304.
\end{description}


\begin{center}
{\large APPENDIX}
\end{center}
\renewcommand{\theequation}{A.\arabic{equation}} \setcounter{equation}{0}
\setcounter{section}{0}

Let $\Sigma_i=\rmcov[X\mid Y=i]$, $\Sigma_{gj}=\rmcov[X\mid Y_g=j]$,
$\Jb=(J_1,\cdots,J_h)^T$ with $J_i=\I(Y=i)$,
$\Jb_g=(J_{g1},\cdots,J_{gs})^T$ with $J_{gj}=\I(Y_g=j)$,
$\rmE[\Jb]=\fb$, and $\rmE[\Jb_g]=\fb_g$. There must exist a code
matrix $G=[G_1,\cdots,G_s]$ with $G_i\in \mathbb{R}^{h}$ containing
only zeros and ones such that $\Jb_g=G^T\Jb$. We may assume $\mu=0$
without loss of generality and, hence, $M=\mb D_\fb \mb^T$ and
$M_g=\mb_gD_{\fb_g}\mb_g^T$. From the definitions of $K_{\rm SIR}$
and $K_{g,\rm SIR}$, we have
$\Gamma=\rmeig(\Sigma^{-1}M;d)$ and
$\Gamma_g=\rmeig(\Sigma^{-1}M_g;d_g)$. Similarly,
$\hat\Gamma=\rmeig(\hat\Sigma^{-1}\hat M;d)$,
$\tilde\Gamma_g=\rmeig(\hat\Sigma^{-1} \hat M_g;d_g)$, and
$\hat\Gamma_g=\rmeig(\hat\Sigma^{-1}\hat P^T\hat M_g\hat P;d_g)$,
where $\hat P=\hat\Gamma\hat\Gamma^T\hat\Sigma$ is an estimator of
$P=\Gamma\Gamma^T\Sigma$ which is the projection matrix onto
$\rmspan(\Gamma)$ relative to the $\Sigma$-inner product.

\begin{proof}[\textbf{Proof of Theorem~\ref{thm.1}}]
By $P^TM_gP=M_g$ and delta method, it suffices to show
$$\Psi=\rmacov\left(U_n^*\right)-\rmacov\left(U_n\right)\geq 0,$$ where
$U_n=n^{1/2}\rmvec(\hat\Sigma^{-1}\hat P^T\hat M_g\hat
P-\Sigma^{-1}M_g)$ and $U_n^*=n^{1/2}\rmvec(\hat\Sigma^{-1}\hat
M_g-\Sigma^{-1}M_g)$.
We first derive the weak convergence of $U_n$. Let
$H_0(M,M_g,\Sigma)= \Sigma^{-1}P^TM_gP$. One has
$H=\partial\rmvec(H_0(M,M_g,\Sigma))/\partial
\rmvec([M,M_g,\Sigma])=[H_1,H_2,H_3]$ by Lemma~4.1 of Tyler (1981),
where $H_1=(I_p\otimes\Sigma^{-1})(I_{p^2}+T_{p,p})\{Q^T\otimes
(M_gM^{+})\}$, $H_2=P^T\otimes (\Sigma^{-1}P^T)$,
$H_3=-(M_g\Sigma^{-1})\otimes\Sigma^{-1}$, $\otimes$ is the
Kronecker product, $T_{p,p}=\sum_{i,j=1}^pE_{ij}\otimes E_{ij}^T$ is
the commutation matrix with $E_{ij}$ being a $p\times p$ matrix with
a one in the $(i,j)$ position and zeroes elsewhere,
$M^{+}$ is the Moore-Penrose inverse of $M$, and $Q=I_p-P$.
From Lemma~1 below and delta method, $U_n=n^{1/2}\rmvec(H_0(\hat
M,\hat M_g,\hat\Sigma)-H_0(M,M_g,\Sigma))$ converges weakly to
$N(0,HWH^T)$, where $W$ is defined in Lemma~1.
As to the weak convergence of $U_n^*$, define $\bar
H_0(M,M_g,\Sigma)= \Sigma^{-1}M_g$ and its differential with respect
to $[M,M_g,\Sigma]$ is calculated to be $\bar H=[0,\bar H_2, H_3]$
with $\bar H_2=(I_p\otimes \Sigma^{-1})$. A similar technique gives
$U_n^*=n^{1/2}\rmvec(\bar H_0(\hat M,\hat M_g,\hat\Sigma)-\bar
H_0(M,M_g,\Sigma))$ which converges weakly to $N(0,\bar HW\bar
H^T)$.

The difference of the asymptotic covariance matrices is $\Psi=\bar
HW\bar H^T-HWH^T=\sum_{i=1}^3\Psi_i$ with $\Psi_1= \bar
H_2W_{22}\bar H_2^T-H_1W_{11}H_1^T-H_2W_{22}H_2^T$, $\Psi_2=\bar H_2
W_{23}H_3^T+H_3 W_{32}\bar H_2^T-H_1
W_{13}H_3^T-H_3W_{31}H_1^T-H_2W_{23}H_3^T-H_3W_{32}H_2^T$, and
$\Psi_3=-H_1W_{12}H_2^T-H_2W_{21}H_1^T$. It is shown in Lemma~2 that
$\Psi_2=0$. Moreover, Lemma~3 implies $\Psi_3=0$. Hence,
$\Psi=\Psi_1$ and we are left to show $\Psi_1\geq 0$. By Lemma~3 and
$Q^T\mb=Q^T\mb_g=0$,
\begin{eqnarray*}
\Psi_1=(I_p\otimes\Sigma^{-1})(I_{p^2}+T_{p,p})\{(M_g-M_gM^{+}M_g)\otimes(Q^T\Sigma
Q)\}(I_{p^2}+T_{p,p})(I_p\otimes\Sigma^{-1}).
\end{eqnarray*}
Since $Q^T\Sigma Q\geq 0$ and is not a zero matrix, it remains to
show $M_g-M_gM^{+}M_g\geq0$. Let $M_g^*=M-M_g$. Since $Y_g$ is a
function of $Y$, $\rmE[X\mid Y_g]=\rmE\{\rmE[X\mid Y]\mid Y_g\}$
and, hence, $M_g^*=\rmE[\rmcov(\rmE[X\mid Y]\mid Y_g)]\geq 0$. It
further implies $M_g-M_gM^{+}M_g=M_g(M_g+M_g^*)^+M_g^*$. By Lemma~4
of Anderson and Duffin (1969), we have $M_g(M_g+M_g^*)^+M_g^*\geq 0$
which proves $\Psi\geq0$. The equality holds if and only if
$M_g(M_g+M_g^*)^+M_g^*=0$, if and only if
$\rmspan(M_g)\bigcap~\rmspan(M_g^*)=\{0\}$ by Lemma~3 of Anderson
and Duffin (1969), if and only if $\rmspan(K_{g,\rm
SIR})\bigcap~\rmspan(K_{\rm SIR}-K_{g,\rm SIR})=\{0\}$.
\end{proof}

\noindent \textbf{Lemma~1.} As $n$ goes to infinity,
$n^{1/2}\rmvec([\hat M, \hat M_g, \hat
\Sigma]-[M,M_g,\Sigma])\stackrel{d}{\rightarrow}N(0,W)\label{lm.1}$,
where the asymptotic covariance matrix $W=[W_{ij}]$, $1\leq i,j\leq
3$, is defined in the proof.
\begin{proof}
The limiting distributions of sample covariance matrix are the same
no matter we know the true mean $\mu=0$ or not.
Thus, we consider $\hat\Sigma=n^{-1}\sum_{i=1}^nX_iX_i^T$ and adopt
a similar strategy of Saracco (1997) to complete the proof.

Let $\ub=(\Jb^T,(\Jb\otimes X)^T, X^T, \Jb_g^T,(\Jb_g\otimes
X)^T,(X\otimes X)^T)^T$, $\rmE[\ub]=\mu_\ub$,
$\Sigma_\ub=\rmcov(\ub)$, and $\bar\ub=\frac{1}{n}\sum_{i=1}^n\ub_i$
with $\ub_i$'s being random copies of $\ub$.
By the central limit theorem we have
$n^{1/2}(\bar\ub-\mu_\ub)\stackrel{d}{\rightarrow}N(0,\Sigma_\ub)$.
Consider
$F_0$ which maps $(a,(b_1,\cdots,b_h),c,d,(e_1,\cdots,e_s),f)$ to
$\rmvec([\sum_{i=1}^ha_i(\frac{b_i}{a_i}-c)(\frac{b_i}{a_i}-c)^T,\sum_{i=1}^sd_i(\frac{e_i}{d_i}-c)(\frac{e_i}{d_i}-c)^T,f~])$
for $a=(a_1,\cdots,a_h)^T\in \mathbb{R}^{h}$, $b_i\in
\mathbb{R}^{p}$, $c\in \mathbb{R}^{p}$, $d=(d_1,\dots,d_s)^T\in
\mathbb{R}^{s}$, $e_j\in \mathbb{R}^{p}$, and $f\in
\mathbb{R}^{p^2}$. By delta method, we deduce that
$n^{1/2}\rmvec([\hat M,\hat
M_g,\hat\Sigma]-[M,M_g,\Sigma])=n^{1/2}(F_0(\bar\ub)-F_0(\mu_\ub))$
converges weakly to $N(0,W)$ with $W=F\Sigma_\ub F^T$, where $F$ is
the differential of $F_0$ at $\mu_\ub$.
A direct calculation then gives
$W_{11}=E_{1}\rmcov(\Jb,\Jb)E_{1}^T+E_{2}\rmdiag(f_{1}^{-1}\Sigma_{1},\cdots,f_{h}^{-1}\Sigma_{h})E_{2}^T$,
$W_{22}=E_{3}\rmcov(\Jb_g,\Jb_g)E_{3}^T+E_{4}\rmdiag(f_{g1}^{-1}\Sigma_{g1},\cdots,f_{gs}^{-1}\Sigma_{gs})E_{4}^T$,
$W_{33}=\rmcov(X\otimes X)$,
$W_{12}=E_1\{\rmcov(\Jb,\Jb_g)(E_3+C_{\mb_g}^*E_4)^T+\rmcov(\Jb,(D_{\fb_g}^{-1}\Jb_g)\otimes
X)E_4^T\}+E_2\rmdiag(\Sigma_1,\cdots,\Sigma_h)\{(GD_{\fb_g}^{-1})\otimes
I_p\}E_4^T$, $W_{23}=E_{3}\rmcov(\Jb_g, X\otimes
X)+E_{4}[\{(D_{\fb_g}^{-1}G^TD_\fb)\otimes I_p\}\Phi-C_{\mb_g}
\rmE\{(D_{\fb_g}^{-1}\Jb_g)\otimes (X\otimes X)^T\}]$, and
$W_{13}=E_{1}\rmcov(\Jb, X\otimes X)+E_{2}(\Phi-C_\mb
\rmE[(D_{\fb}^{-1}\Jb)\otimes (X\otimes X)^T])$, where
$\Phi=\rmE[(D_{\fb}^{-1}\Jb)\otimes \{X(X\otimes X)^T\}]$,
$C_\mb=\rmdiag(m_1,\cdots,m_h)$,
$C_{\mb_g}=\rmdiag(m_{g1},\cdots,m_{gs})$,
$C_{\mb_g}^*=\rmdiag(f_{g1}^{-1}m_{g1},\cdots,f_{gs}^{-1}m_{gs})$,
$E_{1}=[m_1\otimes m_1,\cdots,m_h\otimes m_h]$,
$E_{2}=(I_{p^2}+T_{p,p})\{(\mb D_\fb)\otimes I_p\}$,
$E_{3}=[m_{g1}\otimes m_{g1},\cdots,m_{gs}\otimes m_{gs}]$, and
$E_{4}=(I_{p^2}+T_{p,p})\{(\mb_gD_{\fb_g})\otimes I_p\}$.
\end{proof}

\noindent \textbf{Lemma~2.} Under \textbf{(A1)-(A2)}, $\Psi_2=0$.
\begin{proof}
From $Q^T\mb=Q^T\mb_g=0$ and $Q^T\Sigma P=0$, we have
$\Psi_2=\Psi_{20}+\Psi_{20}^T$ with
$\Psi_{20}=(I_p\otimes\Sigma^{-1})(I_{p^2}+T_{p,p})\{(\mb_gG^TD_\fb-M_gM^{+}\mb
D_\fb)\otimes I_p\}\{(I_h\otimes Q^T)\Phi(P\otimes I_p)\}H_3^T,$ and
it suffices to show $\Psi_{20}=0$. From
$\rmspan(\Gamma)=\mathcal{S}_{Y\mid X}$ and \textbf{(A1)}, we have
$Q^T\rmE[X(X^T\otimes X^T)\mid Y=i](P\otimes
I_p)=\rmE\{(X^TP)\otimes\rmcov(Q^TX\mid \Gamma^TX)\mid
Y=i\}=(m_i^{T}P) \otimes (Q^T\Sigma Q)$ by Lemma~4. It further
implies
$(I_h\otimes Q^T)\Phi(P\otimes I_p)=(\mb^{T}P)\otimes (Q^T\Sigma
Q)$. Substituting this into $\Psi_{20}$ and using
$(\mb_gG^TD_f-M_gM^{+}\mb D_\fb)\mb^{T}=M_g-M_g=0$ to conclude
$\Psi_{20}=0$.
\end{proof}

\noindent \textbf{Lemma~3.} Under \textbf{(A1)-(A2)},
$Q^T\Sigma_iQ=Q^T\Sigma_{gj}Q=Q^T\Sigma Q$ and
$Q^T\Sigma_iP=Q^T\Sigma_{gj}P=0$.
\begin{proof}
Note that $\Sigma_i=\rmE[\rmcov(Q^TX\mid \Gamma^TX)\mid
Y=i]+\rmcov(P^TX\mid Y=i)$ by \textbf{(A1)} and
$\rmspan(\Gamma)=\mathcal{S}_{Y\mid X}$. The result is proved by
Lemma~4. The case of $\Sigma_{gj}$ is similar.
\end{proof}

\noindent \textbf{Lemma~4.} Under \textbf{(A1)-(A2)},
$\rmcov(Q^TX\mid \Gamma^TX)=Q^T\Sigma Q$.
\begin{proof}
From \textbf{(A1)},
$\rmcov(Q^TX\mid \Gamma^TX)=\xi(X)Q^T\Sigma Q \label{lem.5}$
for some positive function $\xi(\cdot)$. Also,
$\Sigma=\rmE[\rmcov(X\mid \Gamma^TX)]+\rmcov(\rmE[X\mid \Gamma^TX])$
implies $Q^T\Sigma Q=\rmE[\rmcov(Q^TX\mid \Gamma^TX)]$. These two
facts gives $\rmE[\xi(X)]=1$. Note that $Q^T\Sigma P=0$ implies
$\rmspan(\Sigma^{1/2}Q)~\bot~\rmspan(\Sigma^{1/2}P)=\mathcal{S}_{Y\mid
Z}$ and, hence, $\rmcov(Q^TX\mid
\Gamma^TX)=\rmcov(Q^T\Sigma^{1/2}Z\mid B^TZ)$ is non-random by
\textbf{(A2)}. Hence, we must have $\xi(\cdot)=1$ which completes
the proof.
\end{proof}

%
%

\newpage

\begin{center}
\centerline{Table~1}\small{\emph{\label{tb.sim} Averages of
Frobenius norms under different $t$ and $(n,p,{\rm CR})$}} for
models~(\ref{ex.2})-(\ref{ex.3}) \vspace{0.1cm}

\tabcolsep=5pt
\begin{tabular}{cccccc}
\hline \hline
Model-(4)  &       &   (100,~10,~$0\%$)    &   (100,~20,~$0\%$)    &   (100,~10,~$25\%$)   &   (50,~10,~$0\%$) \\
\hline
$t_{30}$    &   SIR-SIR &   0.241   &   0.320   &   0.343   &   0.326   \\
    &   SIR &   0.358   &   0.558   &   0.451   &   0.515   \\
\hline
$t_{50}$    &   SIR-SIR &   0.181   &   0.278   &   0.317   &   0.265   \\
    &   SIR &   0.309   &   0.490   &   0.408   &   0.455   \\
\hline
$t_{70}$    &   SIR-SIR &   0.239   &   0.323   &   0.357   &   0.333   \\
    &   SIR &   0.363   &   0.558   &   0.469   &   0.521   \\
\hline \hline
Model-(5)  &       &   (100,~10,~$0\%$)    &   (100,~20,~$0\%$)    &   (100,~10,~$25\%$)   &   (50,~10,~$0\%$) \\
\hline
$t_{45}$    &   SIR-SAVE    &   0.572   &   0.805   &   0.581   &   0.815   \\
    &   SAVE    &   0.676   &   1.042   &   0.697   &   1.002   \\
\hline
$t_{65}$    &   SIR-SAVE    &   1.022   &   1.449   &   1.101   &   1.391   \\
    &   SAVE    &   1.354   &   1.705   &   1.415   &   1.572   \\
\hline
$t_{75}$    &   SIR-SAVE    &   1.129   &   1.600   &   1.365   &   1.538   \\
    &   SAVE    &   1.775   &   2.176   &   1.844   &   1.952   \\
\hline

\end{tabular}
\end{center}


\newpage

\begin{center}
\centerline{Table~2} \small{ \emph{\label{tb.cad} $\hat\Gamma_g$ and
the time-dependent AUC values $\mathcal{A}_t$ and $\mathcal{A}_t^*$
at different time points $t$}} \vspace{0.1cm}

\tabcolsep=5pt
\begin{tabular}{ccccccc}
\hline \hline
$t$ &   CRP &   SAA &   IL-6    &   tHcy    &   $\mathcal{A}_t$ &   $\mathcal{A}_t^*$   \\
\hline
1000    &   -0.400  &   0.580   &   0.465   &   0.643   &   0.748   &   0.760   \\
1500    &   -0.532  &   0.560   &   0.605   &   0.573   &   0.735   &   0.744   \\
2000    &   -0.495  &   0.579   &   0.573   &   0.578   &   0.733   &   0.745   \\
2500    &   -0.619  &   0.529   &   0.690   &   0.531   &   0.693   &   0.708   \\
3000    &   -0.695  &   0.488   &   0.759   &   0.499   &   0.709   &   0.724   \\
3500    &   -0.735  &   0.165   &   0.652   &   0.705   &   0.670   &   0.675   \\
\hline
\end{tabular}
\end{center}

\newpage

\end{document}